\documentclass[prd,amsfonts,amsmath,twocolumn,superscriptaddress,nofootinbib]{revtex4}

\usepackage{color}
\usepackage{graphicx} 
\usepackage{epstopdf}%

\usepackage{mathpazo}      

\newcommand {\be} {\begin{equation}} 
\newcommand {\ba}{\begin{eqnarray}} 
\newcommand {\ee} {\end{equation}} 
\newcommand{\ea} {\end{eqnarray}}

\renewcommand{\epsilon}{\varepsilon}

\preprint{MKPH-T-11-03}

\begin{document}

\title{
Off-axis Excitation of Hydrogen-like Atoms  by Twisted Photons}

\author{Andrei Afanasev}

\affiliation{Department of Physics,
The George Washington University, Washington, DC 20052, USA}

\author{Carl E.\ Carlson}

\affiliation{Department of Physics, The College of William and Mary in Virginia, Williamsburg, VA 23187, USA}

\author{Asmita Mukherjee}

\affiliation{Department of Physics, Indian Institute of Technology Bombay, Powai, 
Mumbai 400076, India}

\date{\today}

\begin{abstract}

We show that the twisted photon states, or photon states with large ($> \hbar$) angular momentum projection ($m_\gamma$)  in the direction of motion, can photoexcite atomic
 levels for a hydrogen-like atom that are novel and distinct  and are not restricted by $m_\gamma$, when   the symmetry axis of the twisted-photon beam does not coincide with the center of the atomic target. Selection rules are given and interesting implications and observables for the above process are pointed out. 

\end{abstract}

\maketitle

\section{Introduction and Motivation}	\label{sec:intro}

The fact that circularly polarized photons carry an angular momentum {$\hbar$} was predicted theoretically and demonstrated experimentally in a seminal experiment by Beth in 1936 \cite{Beth36}. It was also realized \cite{Heitler} (Appendix) that the electromagnetic wave can carry orbital angular momentum in the direction of its propagation if it is constrained in the transverse plane, as in the waveguides. Much later, in 1992, Allen and collaborators suggested \cite{Allen:1992zz} that a special type of light beams that can propagate in vacuum, called Laguerre-Gaussian, predicted as non-plane wave solutions of Maxwell equations, can carry large angular momentum $ J_z \gg \hbar$ associated with their helical wave fronts. At a quantum level such beams can be described in terms of "twisted photons" \cite{molina2007nature}. This concept can be also extended to beams of particles, and electrons in particular \cite{Uchida10}. Detailed reviews of the field are published, {\it c.f.} \cite{Yao11}.  Methods to produce "twisted" light include spiral phase plates, computer-generated holograms \cite{Yao11}, synchrotron radiation in a helical undulator \cite{Sasaki08, Afanasev11}, or in a free-electron laser \cite{Hemsing11}. Theoretical work \cite{Jentschura:2010ap,Jentschura:2011ih} has shown that one can generate twisted photons with high energies of several GeV via Compton back-scattering of laser photons on an energetic electron beam, making such photon beams relevant for nuclear and particle physics. 

An important question is, to what extent absorption of the twisted photons by atoms or nuclei is different from the plane-wave photons? Work by Pic\'on {\it et al.} \cite{Picon10} demonstrated that during photoionization of atoms, the knocked-out electrons carry angular momenta that reproduce the angular momentum of the incoming photons. The references \cite{Picon10,picon2010njp} deal mainly with a special case in which the photon beam's symmetry axis coincides with a center of an atom. In a recent publication \cite{Davis13} the authors analyzed elastic scattering of the twisted photons on a hydrogen atom, again with a restriction that  the atom is located at the center of the optical vortex. We consider a more general case of arbitrary positioned beams and considered photoexcitation of bound states with different quantum numbers in a hydrogen atom.  Such considerations can also be found in~\cite{Jauregui:2004}, which has important observations regarding total angular momentum conservation, although many formulas are given just for the atom on axis case.
After presenting theoretical formalism for excitation of an atom by twisted photons, we point out novel effects caused by large angular momenta of the incoming photons. Our arguments are further corroborated by theoretical calculations showing that a significant fraction of the atomic levels excited by the twisted photons could not be otherwise produced by plane-wave photons.
The angular momentum projection of the twisted photon state is derived in the
appendix.


\section{Basic Formulae}			\label{sec:one}


The twisted photon definition here follows Serbo and Jentschura~\cite{Jentschura:2010ap,Jentschura:2011ih}, although with a more field theory based viewpoint. Another possibility would be to quantize a Laguerre-Gaussian laser mode considered in the original work by Allen et al. \cite{Allen:1992zz} ,
but our main conclusions will not be affected by this choice.

A twisted photon moving in the $z$-direction is 
\begin{align}
\label{eq:twisteddefinition}
| \kappa m_\gamma k_z \Lambda \rangle &= \int \frac{d^2k_\perp}{(2\pi)^2} 
	a_{\kappa m_\gamma}(\vec k_\perp) | \vec k, \Lambda \rangle	\nonumber\\
&= \sqrt{\frac{\kappa}{2\pi}} \  \int \frac{d\phi_k}{2\pi} (-i)^{m_\gamma} e^{im_\gamma\phi_k}  \,
	|\vec k, \Lambda\rangle
\end{align}
where $|\vec k, \Lambda\rangle$ are plane wave states, or momentum eigenstates with fixed longitudinal component $k_z$ and fixed magnitude traverse component,
\be
a_{\kappa m_\gamma}(\vec k_\perp) = (-i)^{m_\gamma} e^{im_\gamma\phi_k} \sqrt{\frac{2\pi}{\kappa}}
	\delta(\kappa - |\vec k_\perp |)	\,.
\ee
The twisted photon state can thus be viewed as a superposition of plane wave states where the momenta form a cone in momentum space with a fixed pitch angle
\be
\theta_k = \arctan\left( \frac{ |\vec k_\perp| }{ k_z } \right)		\,,
\ee
and varying azimuthal angle weighted by a phase $e^{im_\gamma\phi_k}$.

The normalization is
\be
\langle \kappa' m_\gamma' k'_z \Lambda'   | \kappa m_\gamma k_z \Lambda \rangle
	= 2\pi \, 2\omega \delta(k_z-k'_z) \delta(\kappa-\kappa') 
	\delta_{m_\gamma m_\gamma'} \delta_{\Lambda\Lambda'}
\ee
for  $\langle \vec k' \Lambda'  | \vec k \Lambda \rangle 
= (2\pi)^3 2\omega \delta^3(\vec k - \vec k') \delta_{\Lambda\Lambda'}$, and $\omega = |\vec k|$.

Using the photon field operator $A^\mu(x)$,
the wave function of a plane wave photon is
\be
\label{eq:planewave}
\langle 0 | A^\mu(x) | \vec k, \Lambda \rangle 
	= \epsilon^\mu_{\vec k,\Lambda} e^{-ikx}
\ee
and so the wave function of the twisted photon is
\begin{align}
\label{eq:twistedwave}
\mathcal A^\mu_{\kappa m k_z \Lambda}(x)
	&= \langle 0 | A^\mu(x) | \kappa m k_z \Lambda \rangle	\nonumber\\
&= \sqrt{\frac{\kappa}{2\pi}} \  \int \frac{d\phi_k}{2\pi} (-i)^{m_\gamma} e^{im_\gamma\phi_k}  \,
	\epsilon^\mu_{\vec k,\Lambda} e^{-ikx}	\,.
\end{align}
In cylindrical coordinates this is
\begin{align}
\mathcal A^\mu_{\kappa m_\gamma k_z \Lambda}(x)
&= e^{-i(\omega t - k_z z)}	\nonumber\\
& \times
\sqrt{\frac{\kappa}{2\pi}} \  \int \frac{d\phi_k}{2\pi} (-i)^{m_\gamma} e^{im_\gamma\phi_k}  \,
	\epsilon^\mu_{\vec k,\Lambda} e^{i \vec k_\perp{\cdot}\vec x_\perp},
\end{align}
so that the twisted photon in coordinate space has a self-reproducing 2D wave front moving forward at a speed less that the normal speed of light.

The wave front can be given explicitly with help of the Jacobi-Anger formula
\be
e^{i \vec k_\perp{\cdot}\vec x_\perp}
	= \sum_{n= -\infty}^\infty i^n e^{in(\phi_\rho - \phi_k)}  J_n(\kappa\rho)
\ee
where $\phi_\rho$ is the azimuthal angle in coordinate space, $\rho=|\vec x_\perp|$, and $J_n$ is the Bessel function,  and with
\be
\label{eq:epsilonexpand}
\epsilon^\mu_{\vec k \Lambda} = 
	e^{-i\Lambda\phi_k} \cos^2\frac{\theta_k}{2} \eta^\mu_\Lambda
	+ e^{i\Lambda\phi_k} \sin^2\frac{\theta_k}{2} \eta^\mu_{-\Lambda}
	+ \frac{\Lambda}{\sqrt{2}} \sin\theta_k \, \eta^\mu_0
\ee
where the $\eta$'s are constant vectors,
\be
\eta^\mu_{\pm 1} = \frac{1}{\sqrt{2}}  \left( 0,\mp 1,-i,0 \right)	\,,
\quad \eta^\mu_0 =  \left( 0,0,0,1 \right)	\,;
\ee
the photon polarization vector phase is like the Trueman-Wick~\cite{Trueman:1964zzb} phase convention.  Then
\begin{align}
\label{eq:twistedwf}
\mathcal A^\mu_{\kappa m_\gamma k_z \Lambda}(x) &= e^{-i(\omega t - k_z z)}	
\sqrt{\frac{\kappa}{2\pi}}  \,	\Bigg\{
	\frac{\Lambda}{\sqrt{2}} e^{im_\gamma\phi_\rho} \sin\theta_k
	J_{m_\gamma}(\kappa\rho) \, \eta^\mu_0			\nonumber\\[1ex]
& \quad + i^{-\Lambda} e^{i(m_\gamma-\Lambda)\phi_\rho}  \cos^2\frac{\theta_k}{2} 
	J_{m_\gamma-\Lambda}(\kappa\rho) \, \eta^\mu_\Lambda	\nonumber\\[1ex]
& \quad + i^{\Lambda}  e^{i(m_\gamma+\Lambda)\phi_\rho}  \sin^2\frac{\theta_k}{2} 
	J_{m_\gamma+\Lambda}(\kappa\rho) \, \eta^\mu_{-\Lambda}
	\Bigg\}	\,.
\end{align}

As an aside, if we were to write the photon wave function for a plane-wave photon of helicity $\Lambda$, it would be like the above, possibly with some differences of normalization choice, but with pitch angle $\theta_k\to 0$ (including $\kappa\to 0$) and without an azimuthal phase factor, \textit{i.e.}, equivalent to $m_\gamma=\Lambda$ in the preceding equation.

The twisted photon wave front has the feature that the Poynting vector is spiraling forward.  It has azimuthal and $z$ components in cylindrical coordinates, but no radial component.  In detail, the magnetic field for $\Lambda=1$ is 
\begin{align}
B_\rho &=  i\omega \sqrt{\frac{\kappa}{4\pi}} e^{i(k_z z -\omega t + m_\gamma\phi)}
	\nonumber\\
& \quad \times \left( \sin^2\frac{\theta_k}{2} J_{m_\gamma+1}(\kappa\rho) 
		+ \cos^2\frac{\theta_k}{2} J_{m_\gamma-1}(\kappa\rho) \right) \,,	\nonumber\\
B_\phi &=  \omega \sqrt{\frac{\kappa}{4\pi}} e^{i(k_z z -\omega t + m_\gamma\phi)}
	\nonumber\\
& \quad \times \left( \sin^2\frac{\theta_k}{2} J_{m_\gamma+1}(\kappa\rho) 
		- \cos^2\frac{\theta_k}{2} J_{m_\gamma-1}(\kappa\rho) \right) \,,	\nonumber\\
B_z &= \omega \sqrt{\frac{\kappa}{4\pi}} e^{i(k_z z -\omega t + m_\gamma\phi)}
	\sin\theta_k J_{m_\gamma}(\kappa\rho)	\,,
\end{align}
and the electric field is just 90$^\circ$ out of phase with the magnetic field, $\vec E = i \vec B$.  The physical fields are the real parts of the above expressions, and one can see the wave front moves forward at less than the normal speed of light.    Working  physical electric and magnetic fields, the Poynting vector $\vec S = \vec E \times \vec B$ is
\begin{align}
S_\rho &= 0	\,,	\nonumber\\
S_\phi &= \frac{\kappa \omega^2}{4\pi} \sin\theta_k \, J_{m_\gamma}(\kappa\rho)
		\nonumber\\
& \quad \times	\left( \cos^2\frac{\theta_k}{2} J_{m_\gamma-1}(\kappa\rho)
		+ \sin^2\frac{\theta_k}{2} J_{m_\gamma+1}(\kappa\rho)	\right)  \,,	\nonumber\\
S_z &= \frac{\kappa \omega^2}{4\pi}
	\left( \cos^4\frac{\theta_k}{2} J^2_{m_\gamma-1}(\kappa\rho)
		- \sin^4\frac{\theta_k}{2} J^2_{m_\gamma+1}(\kappa\rho)	\right)	\,.
\end{align}

\begin{figure}[htbp]
\begin{center}
\includegraphics[width = 78 mm]{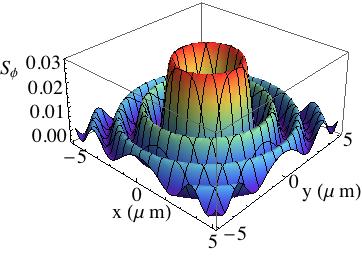}
\caption{The size of Poynting vector azimuthal component as a function of position in the transverse plane.  For this illustration, the photon wavelength is 0.5 microns, the pitch angle is $0.2$ radians, and $m_\gamma=4$. }
\label{fig:sscalar}
\end{center}
\end{figure}

\begin{figure}[htbp]
\begin{center}
\includegraphics[width = 75 mm]{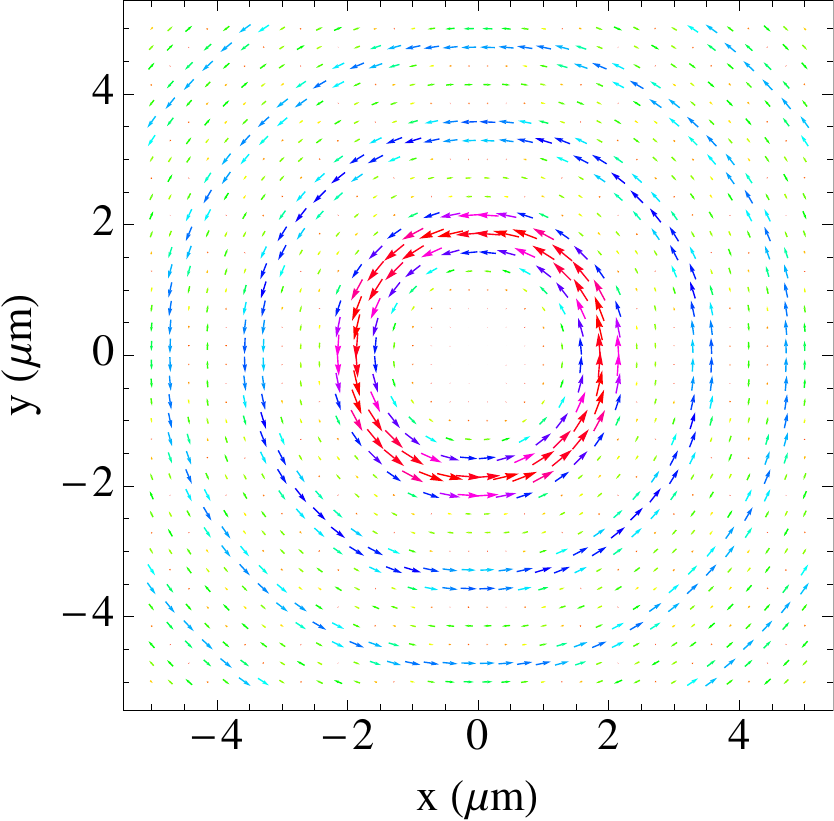}
\caption{A plot of $2\pi\rho$ times $\vec S$ projected onto the transverse plane.  One sees the major bands of $\vec S$ circulating in the same direction, building up the large orbital angular momentum.  Also for this illustration, $\lambda= 0.5 \mu$m, $\theta_k=0.2$ radians, and $m_\gamma=4$.}
\label{fig:svector}
\end{center}
\end{figure}

Figure~\ref{fig:sscalar} shows $S_\phi$ in the transverse plane.  For this illustration, and for the next, the photon wavelength is 0.5 microns, the pitch angle is $0.2$ radians, and $m_\gamma=4$.  The figure shows a bullseye pattern characteristic of twisted photons, with a wide hole in the middle that one can also see from the Bessel functions in the above expressions.  Figure~\ref{fig:svector} shows $\vec S$ in the transverse plane, showing again the bullseye pattern and also showing the circulation of momentum density about the center of the pattern.

The center of the bullseye is currently at the origin in the $x$-$y$ plane.  Shifting it is easily done by applying the translation operator $\exp\{i\hat p{\cdot}b\}$ to the twisted photon state $| \kappa m_\gamma k_z \Lambda \rangle$, where $\hat p^\mu$ is the momentum operator and $b^\mu$ is a constant vector.  There is a \textit{de facto} phase convention in Eq.~(\ref{eq:planewave}), that the momentum eigenstate wave function at $x=0$ is just the polarization vector.  Algebraic effects of the shift are to change the phase in Eq.~(\ref{eq:twistedwave}) from $\exp\{-ikx\}$ to $\exp\{-ik(x-b)\}$ and arguments of the Bessel functions in later equations to $J_\nu(\kappa |\vec x_\perp - \vec b_\perp|)$.


\section{Atomic photoexcitation}			\label{sec:vector}


We will consider excitation by a twisted photon of a hydrogen-like atom from the ground state to an excited state.

In general, the photon's wave front will be traveling in the $z$-direction and the axis of the twisted photon will be displaced from the nucleus of the atomic target by some distance in the $x$-$y$ plane which we will call $\vec b$.  We work out the photoexcitation for this case in this section, and then shall apply the result to two situations.  

\begin{figure}[b]
\begin{center}
\includegraphics{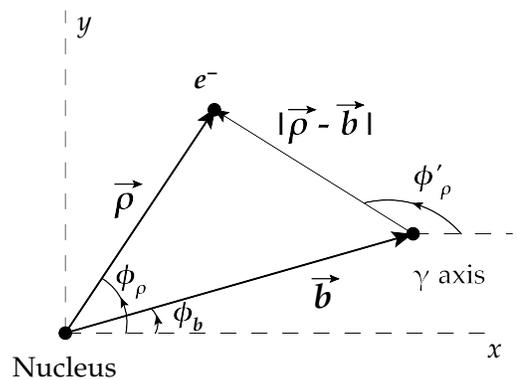}
\caption{Relative positions of atomic state and photon axis, as projected onto the $x$-$y$ plane, with the origin at the nucleus of the atom.}
\label{offaxiscoord}
\end{center}
\end{figure}


One situation will be the case when the twisted photon axis passes directly through the center of the atom's nucleus.  The other will be the case where target atoms are at random locations, and we have to average over all axis to atom separations.

For simplicity, we will treat an atomic state just in terms of its principal quantum number $n_k$, orbital quantum number $l_k$, and magnetic quantum number $m_k$, where $k=i$ for the initial state and $k=f$ for the final state.  

We treat the atom nonrelativistically.  The twisted photon satisfies the Coulomb gauge condition, and the interaction Hamiltonian is
\be
\label{eq:hamiltonian}
H_1 = - \frac{e}{m_e} \vec A \cdot \vec p \,,
\ee
and the transition matrix element is
\begin{align}
S_{fi} &= -i \int dt  \langle n_f l_f m_f | H_1 
	| n_i l_i m_i; \kappa m k_z \Lambda \rangle				\,.
\end{align}

The twisted photon wave function is given in Eq.~(\ref{eq:twistedwf}).

\begin{widetext}

We shall center the atomic nucleus at the origin, with the atomic electron located at $(\rho,\phi_\rho,z)$ in cylindrical coordinates or $(r,\theta_r,\phi_\rho)$ in spherical coordinates.  The twisted photon, moving in the $z$-direction, has its origin in general not centered on the atomic nucleus but displaced to position $\vec b$ in the $x$-$y$ plane.  Relative to the photon axis, the electron position projected onto the $x$-$y$ plane will be at distance $| \vec\rho - \vec b |$ and angle $\phi_\rho'$, as illustrated in Fig.~\ref{offaxiscoord}.

Then 
\begin{align}
S_{fi} &= 2\pi i \delta(E_f-E_i-\omega)  \,\frac{e}{m_e}  \sqrt{\frac{\kappa}{2\pi}}	\,
	\int r^2 dr \, d(\cos\theta_r) \, d\phi_\rho \,  
	R_{n_f l_f}(r) Y^*_{l_f m_f}(\theta_r,0) e^{-i m_f \phi_\rho}	
	\Bigg\{
	\frac{\Lambda}{\sqrt{2}} e^{im_\gamma\phi'_\rho} \sin\theta_k \,
	J_{m_\gamma}(\kappa |\vec \rho - \vec b|) \, \vec\eta_0
											\nonumber\\
&\qquad	+  i^{-\Lambda}\,  e^{i(m_\gamma-\Lambda)\phi'_\rho}  \, \cos^2\frac{\theta_k}{2} \,
	J_{m_\gamma-\Lambda}(\kappa |\vec \rho - \vec b|) \, \vec\eta_\Lambda	
+ i^{\Lambda}  e^{i(m_\gamma+\Lambda)\phi'_\rho}  \,  \sin^2\frac{\theta_k}{2}
	J_{m_\gamma+\Lambda}(\kappa |\vec \rho - \vec b|) \, \vec\eta_{-\Lambda}
	\Bigg\}	e^{i k_z z}		\cdot \vec p \, R_{10}(r) Y_{00}		\,,
\end{align}
where $E_k = E_{n_k}$.

Note that,
\begin{align}
\hat\eta_\lambda {\cdot} \vec p \, R_{10}(r) 
	= -i \hat\eta_\lambda {\cdot} \hat r \, R'_{10}(r) =
	-i \sqrt{\frac{4\pi}{3}} Y_{1\lambda}(\theta_r,\phi_\rho) \, R'_{10}(r)
\end{align}

\noindent The expansion theorem,
\begin{align}
e^{i n \phi_\rho'}  J_n(\kappa |\vec \rho - \vec b|)
	= \sum_{N_1=-\infty}^\infty e^{i N_1 \phi_\rho} e^{-i(N_1-n)\phi_b } 
	J_{N_1}(\kappa\rho) J_{N_1-n}(\kappa b)	\,;
\end{align}
allows us to do the $d\phi_\rho$ integral and obtain
\begin{align}
\label{eq:genresult}
S_{fi} &= -2\pi \delta(E_f-E_i-\omega)  \frac{e}{m_e a_0}  \,  
	\sqrt{\frac{2\pi\kappa}{3}}	\,	e^{i(m_\gamma-m_f)\phi_b} \, J_{m_f-m_\gamma}(\kappa b)
					\nonumber\\
&\times	i^{-\Lambda} \Bigg\{
		 \cos^2\frac{\theta_k}{2} \, g_{n_f l_f m_f \Lambda}	+ 		
		 \frac{i}{\sqrt{2}} \sin\theta_k 	\, g_{n_f l_f m_f 0}
	- 	 \sin^2\frac{\theta_k}{2} \,  g_{n_f l_f m_f, -\Lambda}
	\Bigg\}		\nonumber\\
&\stackrel{\rm def}{=} 2\pi \delta(E_f-E_i-\omega)  \, \mathcal M_{n_f l_f m_f \Lambda}(b)
	\,.
\end{align}

\noindent Note that the energy delta-function requires energy conservation, but in our formalism atomic recoil is neglected, so that overall linear momentum is not conserved. The dimensionless atomic factors are
\begin{align}
\label{eq:reduced}
g_{n_f l_f m_f \lambda} \equiv  - a_0
	\int_0^\infty r^2 dr \ R_{n_f l_f}(r)  \, R'_{10}(r)  
	\int_{-1}^1 d(\cos\theta_r) \, J_{m_f-\lambda}(\kappa \rho) \, 
	Y_{l_f m_f}(\theta_r,0) \, Y_{1 \lambda}(\theta_r,0) e^{i k_z z}		\,,
\end{align} 
and $a_0$ is the Bohr radius.  The quantum numbers of the initial state are tacit, as we always start from the ground state.  As a simple practical matter, 
$-a_0 R'_{10}(r) = R_{10}(r)$, and one also has $\kappa\rho = \omega r \sin\theta_r \sin\theta_k$ and $k_z z = \omega r \cos\theta_r \cos\theta_k$.   Further, one can show that the three terms in the curly bracket above are either all real or else all purely imaginary.

\end{widetext}


\section{On-axis and off-axis atomic excitation}			\label{sec:atomicX}


By way of review, the selection rules for photoexcitation (starting from the ground state) with plane wave photons of helicity $\Lambda$ are, \textit{cf.} 
Refs.~\cite{Schiff},
\begin{align}
m_f &= \Lambda,	\   \
l_f \ge 1,		 \nonumber\\
g^{({\rm pw})}_{n_f l_f, m_f = \Lambda, \Lambda} 
	&\propto \left( \omega a_0 \right)^{l_f - 1}  ,
\end{align}
where $g^{({\rm pw})}$ is the plane wave analog of the reduced atomic amplitudes shown in Eq.~(\ref{eq:reduced}), and shows the suppression that follows when higher photon partial waves are needed.

From Eq. (\ref{eq:genresult}) it can be seen that the magnitude of the result depends on 
$m_\gamma$ only through the argument of the Bessel function $J_{m_f-m_\gamma}(\kappa b)$.  When a twisted photon strikes an atom centered on its axis, the impact parameter $b=0$ and we immediately obtain $m_f = m_\gamma$ from the Bessel function in the general result,
Eq.~(\ref{eq:genresult}),
\begin{equation}
J_{m_f-m_\gamma}(\kappa b) \to J_{m_f-m_\gamma}(0)= \delta_{m_f m_\gamma}	\,.
\end{equation}
That is, the only final states that can be produced are those that can absorb the full projected angular momentum of the twisted photon.

\begin{figure}[h]
\begin{center}
\includegraphics[width = 74 mm]{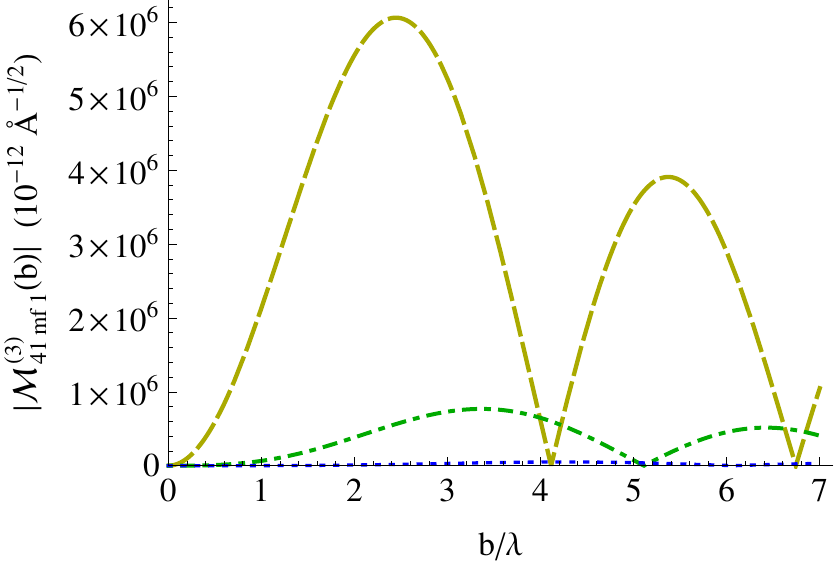}
\includegraphics[width = 74 mm]{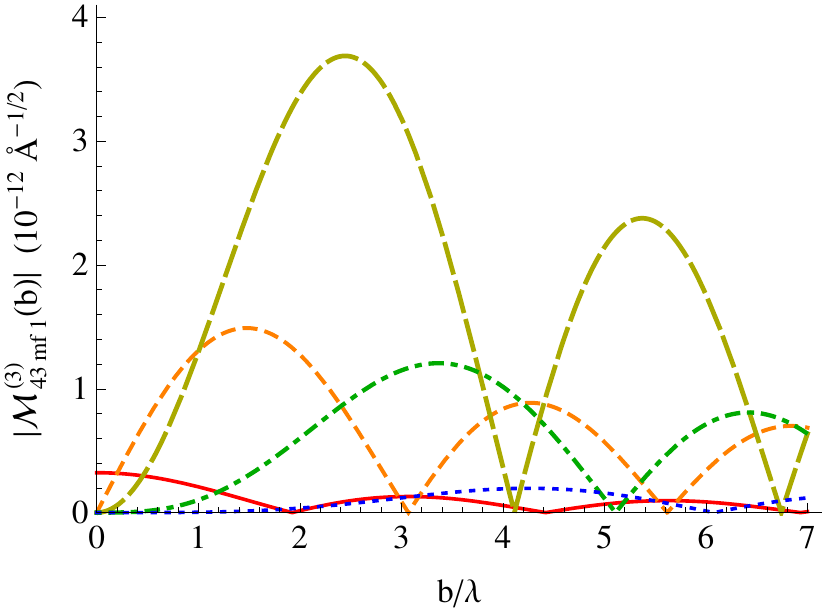}

\caption{Size of the transition amplitude $\left|  \mathcal M^{(m_\gamma=3)}_{n_f l_f m_f \Lambda}(b)  \right|$ for particular quantum numbers $n_f, l_f,\Lambda$ and several $m_f$; here the photons are circularly polarized.  Upper graph is for the final state $n_f=4, \ l_f=1$; the state $m_f=1$ is allowed by electric-dipole selection rules for plane waves, while $m_f=0,-1$ are unique for the twisted photons. Lower graph is for the final state $n_f=4, \ l_f=3$. The curve styles for both graphs are:  $m_f=3$ is the red solid curve, $m_f=2$ is orange and medium dashed, $m_f=1$ is gold and long dashed, $m_f=0$ is green and dot-dashed, $m_f=-1$ is blue and dotted, and transitions to other $m_f$ are quite small and not plotted. }
\label{fig:impact}
\end{center}
\end{figure}

However, the atom does not have to be far off the photon axis before other amplitudes, 
not satisfying the above selection rule, play an important role.  This was also observed 
the in simulations done in \cite{picon2010njp}. In our numerical calculations as well as 
in the plots, unless otherwise stated, we take a twisted photon state of wavelength 97.2 
nm (set by the Hydrogen atom spacing), $m_\gamma=3$, $\Lambda=1$ and $\theta_k=0.2$ radians. 
  As an illustration, we plot in Fig.~\ref{fig:impact} the amplitudes 
$\left|  \mathcal M_{n_f l_f m_f \Lambda}(b)  \right|$ for the example 
of $n_f=4$, $l_f = 1$  and photon angular momentum along the direction of 
motion $m_\gamma=3$ (upper plot) and $n_f=4$, $l_f = 3$, (lower plot). 
Note the relative strength of the amplitudes is much higher, by about six
 orders of magnitudes, for the transition into $l_f=1$ state vs $l_f=3$, in 
accordance with the selection rules presented below.

In Fig.~ \ref{fig:impact} the horizontal axis is the impact parameter $b$ 
in units of the photon wavelength $\lambda$.  Already with an impact 
parameter of less than half a wavelength, amplitudes that do not satisfy
 the $m_f = m_\gamma$ selection rule are becoming important.  The amplitude 
with the largest peak is the one with $m_f=1$, which is the only amplitude 
one would have with a plane-wave photon polarized with helicity $\Lambda = 1$.
   In a near-field setup, when the twisted light outside 
of the central (small-$b$) areas is blocked by a screen, one can study 
whether the enhancement of the transition to $m_f=m_\gamma$ for the photons 
with large $m_\gamma$, as seen in Fig.~\ref{fig:impact}  (lower plot, 
continuous curve), would lead to enhanced attenuation of these photons by 
atoms, making matter more opaque to them.

The general selection rules for the off-axis case are
\begin{align}
m_f &= {\rm any},	\  \	
l_f \ge \left| m_f \right|	,	
\nonumber\\
g_{n_f l_f, m_f, \Lambda} &\propto \left( \omega a_0 \cos\theta_k \right)^{l_f - |m_f - \Lambda| - 1}
	\left(\omega a_0 \sin\theta_k \right)^{ |m_f - \Lambda|}		,
\end{align}
where in the last line, the first factor is absent if its exponent is negative.

Note that  we neglect atomic recoil. For manifest conservation of both linear and angular momentum, the atom's recoil momentum can be taken into account~\cite{Jauregui:2004}.


\section{Cross sections for randomly distributed targets}	\label{sec:xsctn}



\subsection{Cross section calculation}			\label{sec:xsc}


 In general,  
the location of the photon axis cannot be controlled at the level of the photon wavelength, 
we should average over the transverse 
separations.  For transverse separation $\vec b$,

\begin{equation}
\sigma^{(m_\gamma)}_{n_f l_f m_f \Lambda} = 2 \pi \delta(E_f -E_i-\omega_\gamma)
 \frac{ |\mathcal M^{(m_\gamma)}_{n_f l_f m_f \Lambda}(b)|^2 }{f} \,.
\end{equation}
where $f$ is the incoming flux and a suitable sum or average over spins is implied.  

With the target atoms uniformly distributed, the cross section averaged over atom location is
\begin{align}
&\overline\sigma_{n_f l_f m_f \Lambda} 
	=  \frac{ 2 \pi \delta(E_f -E_i-\omega) }{f}
					\nonumber\\
&\hskip 3.8 em	\times   \frac{1}{\pi R^2} \int d^2b \ 
		|\mathcal M_{n_f l_f m_f \Lambda}(b)|^2	
					\nonumber\\[1.2ex]
&= \frac{ 2 \pi \delta(E_f -E_i-\omega) }{f}
	\frac{1}{\pi R^2}	 \frac{ 2\pi e^2 \kappa }{3 m_e^2 a_0^2}
											\nonumber\\
&\,\times	\int_0^R 2\pi b \, db \  	J^2_{m_f-m_\gamma}(\kappa b)
\times	
	\Big|
	\cos^2 \frac{\theta_k}{2} \,	g_{n_f l_f m_f \Lambda}
											\nonumber\\
&\quad	+ \frac{i}{\sqrt{2}} \sin\theta_k 	\, g_{n_f l_f m_f 0}
	-	\sin^2\frac{\theta_k}{2} \,	g_{n_f l_f m_f, -\Lambda}  \Big|^2
\,.
\end{align}

\noindent Outside the integration measure, the only $b$ dependence is in the Bessel function, and the useful integral is
\begin{equation}
\lim_{R\to\infty} \int_0^R  b \, db \ J^2_{m_f-m_\gamma}(\kappa b) 
	= \frac{R}{\pi \kappa}		\,,
\end{equation}
independent of index.

For the flux we take the average density of the twisted photon state times the incoming wave front velocity $k_z/\omega$.  The target is unit normalized.  The density of the twisted photon state, with our normalization and averaged over a disk of radius $R$, can be worked out and leads to
\begin{equation}
f = \rho_{\rm avg}\frac{k_z}{\omega} = \frac{2 k_z}{\pi^2 R}		\,.
\end{equation}

Thus
\begin{align}
\overline\sigma_{n_f l_f m_f \Lambda} &= 2\pi \delta(E_f -E_i-\omega)
	\frac{ 8\pi^3 \alpha^3  }{3  k_z} \  \bigg|
	\cos^2 \frac{\theta_k}{2} \,	g_{n_f l_f m_f \Lambda}
											\nonumber\\
&\quad	+ \frac{i}{\sqrt{2}} \sin\theta_k 	\, g_{n_f l_f m_f 0}
	-	\sin^2\frac{\theta_k}{2} \,	g_{n_f l_f m_f, -\Lambda}  \bigg|^2
\,.		
\end{align}


\subsection{Unique twisted photon features}			\label{sec:tpf}


For the twisted photon centered on target, there is the dramatic result that the magnetic quantum number of the final atomic state must equal the corresponding $z$-projection of the angular momentum of the twisted photon.   This constraint is relaxed tor the general case of random target location, but there are still features unique to twisted photons.

Photoexcitation, starting from the ground state, by a plane-wave photon of a certain helicity leads only to final states whose magnetic quantum number equals the helicity. Twisted photons, on the other hand, photoexcite states with a large range of magnetic quantum numbers $m_f$.  Values of $m_f$ impossible for plane-wave photons are produced even when the twisted photons enter a medium with random target locations.

Twisted photons also produce the $m_f=\Lambda$ states that plane-wave photons necessarily lead to.  But the interest is in the $m_f \ne \Lambda$ states unique to twisted photon production.  To quantify the probability of finding these states, we define a ratio for a fixed $\Lambda$ which compares the rate for producing final states that are unique to twisted photons to the total rate where the twisted photon produces all final states, including $m_f = \Lambda$, for a given energy level characterized by quantum numbers $(n_f,l_f)$ (and for the case of a large interaction region with random target locations),
\begin{align}
f_{\rm twisted} = 
	\frac{\sum_{\stackrel{\scriptstyle{m_f = - l_f,}}{m_f\ne \Lambda}}^{m_f=l_f}
	\overline\sigma_{n_f l_f m_f \Lambda}}
	{\sum_{m_f = - l_f}^{m_f=l_f}
	\overline\sigma_{n_f l_f m_f \Lambda}}	  \,.
\end{align}

The "twisted photon ratio," $f_{\rm twisted}$, evaluates what fraction of the final states excited by the twisted photon could not have been produced by a plane-wave photon.  As a numerical example, we evaluate this ratio for final states with varied values of $n_f$ and  $l_f$ and the result is  
\begin{align}
f_{\rm twisted}\left[ {\rm gnd.\ state} \to (n_f=4, l_f=1) \right] = 2.0\%,	\nonumber\\
f_{\rm twisted}\left[ {\rm gnd.\ state} \to (n_f=4, l_f=3) \right] = 20.3\%.	
\end{align}

A comparison between the total photoproduction rate from twisted photons and from plane-wave photons is,
\begin{equation}
r_{\rm twisted} = \frac{ \sum_{m_f = - l_f}^{m_f=l_f}
	\overline\sigma_{n_f l_f m_f \Lambda} } 
	{\sigma_{n_f l_f \Lambda \Lambda}^{(pw)}}	\,.
\end{equation}
For the above-chosen final states the other ratio works out to $r_{\rm twisted} = 1.02$.

The ratios $f_{twisted}$ and $r_{twisted}$ can be measured in experiments. They can provide a tool to identify twisted photons arriving from point-like sources, no matter if they come from distant stars or produced in the lab. In particular, presence of $m_f=0$ state in the atomic excitation produced by the photons coming from a well-defined direction would be a definitive signal of a twisted photon absorption.

Let us discuss one more special feature of the twisted photons and compare the probabilities of photoexcitation of an (unpolarized) atom by the photons with opposite helicities $\pm\Lambda$. For plane-wave photons these probabilities are identical due to parity conservation. For twisted photons with a fixed $z$-projection of orbital angular momentum but opposite helicities such an asymmetry would not violate parity, because the corresponding photon states do not transform into each other via parity transformation. The corresponding helicity asymmetry $A_{\Lambda}$ can be defined as
\begin{equation}
A_\Lambda(n_f,l_f,\overline{m_\gamma})=\frac{ \sum_{m_f = - l_f}^{m_f=l_f}
	(\sigma^{(\overline{m_\gamma}-1)}_{n_f l_f m_f \Lambda=-1} -\sigma^{(\overline{m_\gamma}+1)}_{n_f l_f m_f \Lambda=1} )} 
	{\sum_{m_f = - l_f}^{m_f=l_f}
	(\sigma^{(\overline{m_\gamma}-1)}_{n_f l_f m_f \Lambda=-1} +\sigma^{(\overline{m_\gamma}+1)}_{n_f l_f m_f \Lambda=1} )}
\end{equation}
  \begin{figure}[t]
\begin{center}
\includegraphics[width = 74 mm]{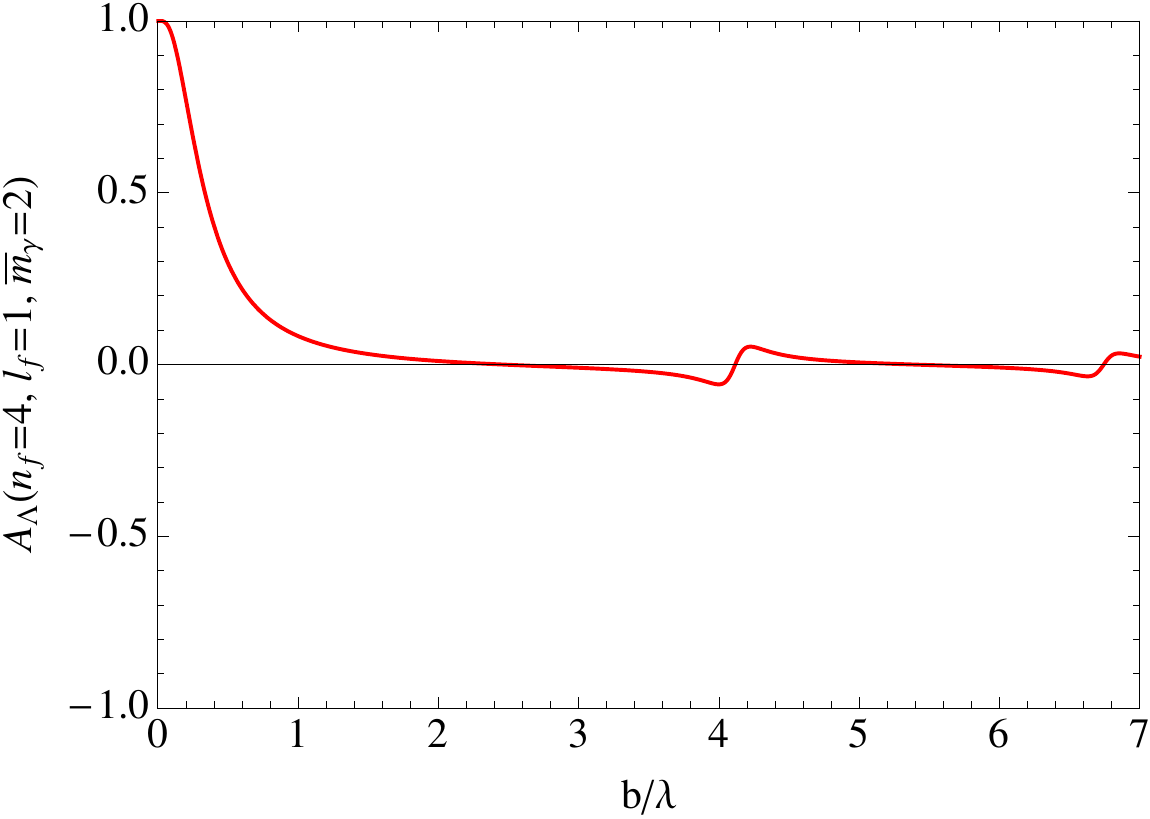}
\caption{Helicity asymmetry $A_{\Lambda}$ for excitation of the atomic state $n_f=4$, $l_f=1$ by a twisted photon with $\overline{m}_\gamma$=2.}
\label{fig:APV}
\end{center}
\end{figure}
The value of $\overline{m_\gamma}$ is zero for plane-wave photons; for the twisted photons it is controlled by the method of their generation, and in paraxial approximation it would correspond to $z$-projection of the orbital angular momentum. The asymmetry $A_\Lambda$ turns to zero after averaging up to infinite values of the impact parameter $b$; it is large for the central areas of the optical vortex, as shown in Fig.~\ref{fig:APV}.
This interesting observable can be studied experimentally for micro-particles placed in the inner area of the optical vortex; it indicates that the electromagnetic fields due to spin and orbital angular momentum of the twisted photons add up coherently, leading to distinctively different strength of interaction at a given transverse-plane location if the direction of spin is flipped. The corresponding figure-of-merit $A_\Lambda^2\sigma$, is small at b=0, but peaks near $b/\lambda$=0.6, where $A_\Lambda$ is about 20\%.


\section{Summary and Discussion}			\label{sec:disc}


In this paper we developed a formalism for photoexcitation of an atom with a beam of twisted photons, using a hydrogen atom as an example. In the derivation, we use an expansion \cite{Jentschura:2010ap,Jentschura:2011ih} of the twisted photon states in terms of plane waves. We show that in a special case when the photon beam axis coincides with the atomic center, the transitions between atomic levels obey angular momentum selection rules, similar to the conclusions made in Ref.\cite{Picon10} for photoionization, resulting in an excited state with a magnetic quantum number ($J_z=m_1$) exactly matching $m_\gamma$ of the incoming beam. We also recover standard electric-dipole angular momentum selection rules \cite{Schiff, Davydov} in the limit of plane-wave photons. 

Next, we extended our calculation to a more general case of the atoms located away from the photon beam axis and analyzed the amplitudes of various transitions as a function of the beam center position in units of photon wavelength, $b/\lambda$. In this case the magnetic quantum number of the photoexcited state no longer matches $m_\gamma$, and a range of final-state quantum numbers is generated. Relative magnitudes of various transitions were studied both analytically and numerically. It was found that after we average over the beam position, the amplitude allowed by standard plane-wave dipole selection rules quickly gains strength and makes a dominant contribution to photoabsorption. 

In order to quantify the role of photon orbital angular momentum, we introduced several observables that describe probabilities of excitation of the $m_f\neq \Lambda$ states that are forbidden to plane waves. The important finding is that relative probability of transitions to "forbidden" states can reach tens of per cent in the examples we considered ($\lambda=$100 nm, pitch angle= 0.2 rad, m=3). This is in stark contrast with familiar selection rules \cite{Schiff, Davydov} , according to which the probability of absorption of photons with higher multipolarity is suppressed by powers of $(ka_0)^2$ that corresponds to about six orders of magnitude for visible light.

Given a noticeable effect arising from the orbital angular momentum of the photons, our predictions can be verified experimentally.









\begin{acknowledgments}

CEC thanks the National Science Foundation for support under Grants
 PHY-0855618 and PHY-1205905. Work of AA was supported by The George 
Washington University. AM thanks Jefferson Lab and the College of William
and Mary for hospitality during the initial stages of this work. 

\end{acknowledgments}

\section{Appendix} \label{sec: appendix}

\subsection{Angular Momentum Projection}			


The factor $\exp\{im_\gamma\phi_k\}$ in the twisted photon state should give the state a $z$-component of angular momentum that is at least approximately $m$.  This is stated in a number of sources, but we have not seen a claim of exactness.   In fact we can prove that the total angular momentum projected in the longitudinal direction is exactly $m_\gamma$, at least in the sense of expectation values.

From the Noether curent corresonding to rotations, one gets the angular momenta.  The result can be found, for example, in Bjorken and Drell \cite{Bjorken-Drell}, Eq.~(14.22), and is
\be
J^{ij} = \epsilon^{ijk} J^k = \int d^3x  
: \dot{\vec A} \cdot (x^i \partial^j - x^j \partial^i) \vec A - (\dot A^i A^j - \dot A^j A^i) :
\ee
We will speak of the first term as the orbital angular momentum and the second as the spin.  

Regarding the spin term, one can consider a direct calculation with the usual expansion and commutation relation in terms of plane wave states,
\be
A^\mu(x) = \sum_\lambda \int \frac{d^3q}{(2\pi)^3 2\omega_q}
	\left( a_{\vec q \lambda} \epsilon^\mu_{\vec q \lambda} e^{-iqx} + 
	a^\dagger_{\vec q \lambda} \epsilon^{\mu *}_{\vec q \lambda} e^{iqx} \right)	\,,
\ee
and
\be
\big[ a_{\vec q \lambda}, a^\dagger_{\vec k \Lambda} \big] = 
	(2\pi)^3 2\omega \delta^3(\vec q - \vec k) \delta_{\lambda\Lambda}	\,.
\ee

After noting that the $a^\dagger a^\dagger$ and $aa$ terms in $J^3$(spin) do not contribute to the matrix element 
below, one can show the $a^\dagger a$ terms lead to
\begin{align}
\langle k'\Lambda'| J^3({\rm spin}) |k\Lambda \rangle
	= 2i\omega (2\pi)^3 \delta^3(\vec k - \vec k') 
\big( \vec\epsilon_{\vec k \Lambda} \times \vec\epsilon^*_{\vec k \Lambda'} \big)^z  .
\end{align}
After showing
\be
\big( \vec\epsilon_{\vec k \Lambda} \times \vec\epsilon^*_{\vec k \Lambda'} \big)^z
	= -i \Lambda \cos\theta_k \delta_{\Lambda\Lambda'}	\,,
\ee
this becomes
\begin{align}
\langle k'\Lambda'| J^3({\rm spin}) |k\Lambda \rangle
	= \Lambda \cos\theta_k  \ \langle k'\Lambda' |k\Lambda \rangle .
\end{align}
There is no $\phi_k$ dependence above, and since the twisted photon states each have fixed $\Lambda$ and $\theta_k$, one can promote the states to twisted photon states and obtain,
\be
\frac{
\langle \kappa' m_\gamma k'_z \Lambda | J^3({\rm spin})    | \kappa m_\gamma k_z \Lambda \rangle}
{
\langle \kappa' m_\gamma k'_z \Lambda  | \kappa m_\gamma k_z \Lambda \rangle}
	= \Lambda \cos\theta_k 	\,.
\ee

Continuing to the orbital angular momentum (OAM) piece, we need
\begin{align}
&\langle \kappa' m_\gamma k'_z \Lambda | J^3({\rm OAM})  | \kappa m_\gamma k_z \Lambda \rangle
		\nonumber\\
& \quad = \langle \kappa' m_\gamma k'_z \Lambda | 
	\int d^3x : \dot{\vec A} \cdot \frac{\partial \vec A}{\partial \phi_\rho} :
	| \kappa m_\gamma k_z \Lambda \rangle
\end{align}
We pursue a different calculation here, still noting that within the normal ordering terms with two creation or two annihilation operators give zero.  For contributions where an $a_{k \lambda}$ comes from
$\partial \vec A/\partial \phi_\rho$ and an $a^\dagger_{k' \lambda'}$ comes from 
$\dot{\vec A}$, the result is unchanged by inserting a vacuum intermediate state. The same is true for the reverse contribution.  Hence
\begin{align}
&\langle \kappa' m_\gamma k'_z \Lambda | J^3({\rm OAM})  | \kappa m_\gamma k_z \Lambda \rangle
		\nonumber\\
& \quad = 2 \int d^3x  \ 
	\langle \kappa' m_\gamma k'_z \Lambda |  \dot{\vec A} |0 \rangle \cdot 
	\langle 0| \frac{\partial \vec A}{\partial \phi_\rho}
	| \kappa m_\gamma k_z \Lambda \rangle	\,.
\end{align}
We can use the results for the twisted state wave functions, Eq.~(\ref{eq:twistedwf}),  and known Bessel function integrals to obtain
\begin{align}
&\frac{
\langle \kappa' m_\gamma k'_z \Lambda | J^3({\rm OAM})    | \kappa m_\gamma k_z \Lambda \rangle}
{
\langle \kappa' m_\gamma k'_z \Lambda  | \kappa m_\gamma k_z \Lambda \rangle} \nonumber\\
	& \quad = \frac{1}{2} m_\gamma \sin^2\theta_k + (m_\gamma-\Lambda) \cos^4\frac{\theta_k}{2}
	+ (m_\gamma+\Lambda) \sin^4\frac{\theta_k}{2}		\nonumber\\
	&\quad = m_\gamma - \Lambda\cos\theta_k	\,.
\end{align}

Combining the results,
\be
\frac{
\langle \kappa' m_\gamma k'_z \Lambda | J^3    | \kappa m_\gamma k_z \Lambda \rangle}
{
\langle \kappa' m_\gamma k'_z \Lambda  | \kappa m_\gamma k_z \Lambda \rangle}	=	m_\gamma	\,.
\ee
The total angular momentum projection in the direction of motion is precisely $m_\gamma$. The value of $m_\gamma$ can be controlled in the lab by the means the beam of twisted photons is generated, for example, by the use of spiral phase plates or computer-generated holograms \cite{Yao11}.


\bibliography{TwistedPhoton}

\end{document}